# The stabilization of width of multiple-quantum NMR spectrum induced by decoherence

A.A. Lundin[1], V.E. Zobov[2]

N.N. Semenov Institute of Chemical Physics, RAS, 117977, Moscow, Russia,

L.V. Kirensky Institute of Physics, Siberian Branch, RAS, 660036, Krasnoyarsk, Russia,

e-mail: 1) andylun@orc.ru; 2) rsa@iph.krasn.ru.

**Abstract**. As opposed to traditional methods of multiple-quantum NMR spectroscopy authors of the article [G.A. Alvarez, D. Suter, Phys. Rev. A **84**, 012320 (2011)] generated the effective double-quantum Hamiltonian, with the slight adding usual secular dipole-dipole Hamiltonian to the first one at the stage of the correlations preparing (so the perturbation appeared). At this framework the width of MQ spectra reveals tending to any constant amount with the growing up of the time. Also there were shown that the spectral width decreases with increasing perturbation strength. The growing up of the cluster size as it was supposed by G.A. Alvarez and D. Suter was restricted by mentioned perturbation. We are assuming an alternative explanation. The growing up of cluster size still goes on but the width of MQ spectrum becomes stable because of different decay rate of MQ-spectral components in dependence on theirs location in spectrum. Here we are calculating the widths of MQ spectra in dependence on the time of "preparation" for different values of the perturbation strength and by this way we get the dependence of stabilized amount of MQ-spectrum width on mentioned strength. So, the excellent compliance obtains with the analogues experimental relation observed in Ref. [Phys. Rev. A **84**, 012320 (2011)].

## 1. INTRODUCTION

Multiple-quantum (MQ) solid-state nuclear magnetic resonance (NMR) represents a power tool for precise control of nuclear spin dynamics allowing the performance of experimental studies of many-body systems interacting through natural and specifically designed couplings [1-3]. At the same time, the states that arise in MQ spectroscopy and develop under the action of certain radio frequency (rf) pulse sequences on the spin system and which are called multiple spin coherences (or MQ coherences depending on the experimental conditions) are described by multiparticle time correlation functions (TCF)s of a complex structure. These coherences and their dynamics provide a powerful and sometimes irreplaceable possibilities for studying particles behavior in different systems such as their clustering, origination of local structures located, for example, on surfaces or in liquid crystals, in nanocavities, and so on [4-6].

We also should note that the processes of origination and decay of multi-particle spin correlations are very important for studying of the contemporary methods of quantum information processing and quantum computations [7]. When being realized, the coherences prepared in a nuclear spin subsystem are controlled by rf pulses initiating the behavior of the required processes. At the same time the methods of MQ spectroscopy allow the experimental study processes of growing up of multi-particle (multispin) correlations in the spin system and that is unique feature of above methods [8-10]. Let's mark also that a thorough monitoring of a quantum register (described by higher-order TCFs) is required for practical realization of the vast potential of quantum computers, and the thoroughness of this monitoring must increase with the number of qubits (spins) contained in the system because the "fragility" of the arising clusters increases with the number of correlated spins as the consequence of the increasing intensity of different relaxation processes destroying the quantum mechanical superposition of states in the system.

Indeed, the exponential growth [11, 12] the number of spins implemented in dynamical correlations as it has been experimentally demonstrated [8-10, 13-15] actually means that in a time about 0.01 sec all the crystal becomes dynamically correlated cluster: $exp(a_0 t) \sim 10^{23}$, if for example $a_0 \sim 0{,}01 \mu s^{-1}$. The above number conforms to appropriate constants [12, 16] for adamantane and for $CaF_2$ with magnetic field along [111]. But in spite of rather correct theoretical explanation of exponential growth of correlation [11] in ideal system the exponential growth cannot be being for all the time in reality. As we mentioned earlier there are a lot of different relaxation processes concerning with spin-lattice relaxation non-ideal apparatus, and so on. Any case here and what follows below in the present paper we are not going to discuss all possible causes invoke the failure of exponential growth of correlations but would concentrate on clarification of experimental results published in articles [13-15].

One of the possible chances to explain the eventual termination of exponential growing up of correlations was investigated in articles [13-15]. In these articles authors supposed the existence of Anderson-like localization of dynamical correlations in nuclear spin system if one slightly perturbed [13-15, 17]. But as we could guess the idea of localization what offered in mentioned articles for the explanation of the observed results perhaps is not too correct. One of the most substantial justification for our conjecture is the absence a significant physical arguments for localization at explored systems. Particularly the most of significant reasons for the absence of localization there are high temperature experimental regimen and the lack of inhomogeneous broadening of NMR spectrum.

Quite recently, we assumed [18] that the main reason for the failure of expected exponential growth of cluster is the consequence of experimental method used in [13-15]. Actually correlated cluster continue growing up in exponential manner, but the growth masks by the method of monitoring process. And namely the width of MQ spectrum becomes stable because of different decay rate of MQ-spectral components in dependence on theirs location in spectrum. In the present article, we are going to demonstrate the explicit explanations of results [13-15] based on developing of our preliminary guess [18]. Results of our theory coincide well with the experimental results.

## 2. GROWTH AND DECAY OF MQ COHERENCES IN SEPARATE PROCESSES OF FORMATION AND DEGRADATION OF CORRELATIONS IN A SPIN SYSTEM

Usually in the conventional MQ NMR spectroscopy, the original Hamiltonian of the secular part of internuclear dipole–dipole interactions (DDI) coupling nuclei in a crystal [19]

$$H^{dd} = H_d = \sum_{i \neq j} b_{ij} S_{zi} S_{zj} - (1/2) \sum_{i \neq j} b_{ij} S_{+i} S_{-j} \qquad (1)$$

is transformed into the effective Hamiltonian by sequence of rf pulses (here $b_{ij}$ are DDI constants [19] and $S_{\pm j} = S_{xj} \pm i S_{yj}$):

$$H^{\pm 2} = H_0 = (-1/4) \sum_{i \neq j} b_{ij} (S_{+i} S_{+j} + S_{-i} S_{-j}) . \qquad (2)$$

The last one is nonsecular with respect to the external magnetic field. Under the effect of this Hamiltonian during the so-called "preparation period" of length $T = N_0 \tau_0$ (or $T = N(\tau_0 + \tau_1)$, see Fig.1), the original magnetization is transformed into various rather complicated TCFs what depend on the product of a various number of spin operators. The density matrix defining this state is

$$\rho(T) = \exp\{iH_{eff}T\}\rho_{eq}\exp\{-iH_{eff}T\} = \sum_M \rho_M(T), \qquad (H_{eff} = H^{\pm 2}), \qquad (3)$$

and it conveniently can be represented as a sum ($\rho_M$) of off diagonal elements with a certain difference of $M$ magnetic quantum numbers, called MQ coherences (also $M$ is the order of a coherence and this number simultaneously enumerates the position of the coherence in the MQ spectrum). The coherences are marked by a phase shift $\varphi$ [1-3,8-10,13-15]. The arising phase shift is proportional to $M\varphi$. At the next stage, the intensity of an $M$-quantum coherence can be separated as the $M$th harmonic of an appropriate Fourier series.

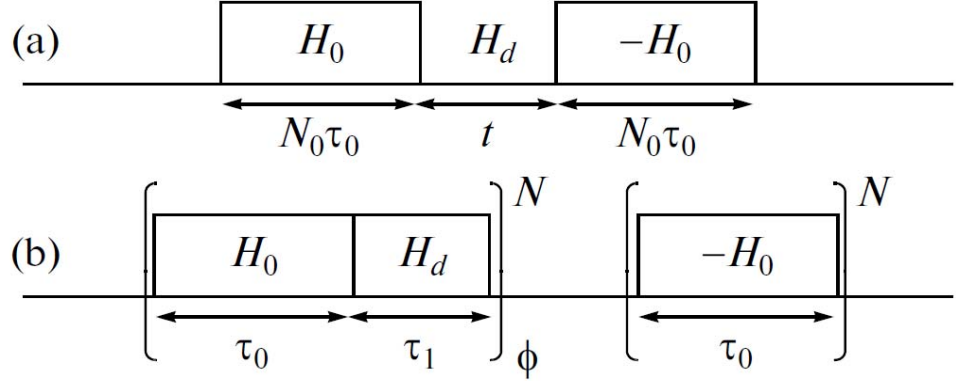

Fig. 1. The experimental schemes implemented in the articles [8- 10] (a) and [13-15] (b).

Following the simplest phenomenological statistical model [1], let's assume that the distribution of coherences of different orders in the MQ experimental spectra can be described by Gaussian:

$$g_M(T) \propto \exp\left\{-\frac{M^2}{K(T)}\right\}. \qquad (4)$$

The dispersion ($K(T)/2$) the above distribution in this model is determined by the number $K(T)$ spins among which a dynamic correlation is established due to interaction Eq. (2) over the preparation time $T$. This number, called the number of correlated spins, or the effective cluster size, increases with the preparation time. The experimental results for the effective cluster size in adamantane from the Refs. [8, 9] are shown at Fig. 2. In strict compliance with our theory from Refs. [11-12] the results demonstrate exponential grows of cluster size:

$$K(T) = \exp(a_0 T). \qquad (5)$$

The suitable value of above parameter is: $a_0$=0.0083 (1/µs).

Further, according to the experimental scheme Fig.1(a) (see Ref. [8]), these coherences are relaxed over a period of time $t$ under the action of the DDI from Eq. (1) as [20]:

$$\Gamma_{0M}(t) = \exp\{-A^2 M^2 t^2\}\exp\{-Kb^2 t^2/2\}. \qquad (6)$$

The Gaussian function in Eq. (6), describes the decay of an MQ coherence as a function of the order $M$. It was obtained in [20] under the condition that each spin in the lattice is surrounded by a large number of approximately equivalent neighbors.

After the period of free evolution, a new sequence of pulses is applied to the system, that reverses the sign of the effective Hamiltonian (2); thus, a time reversal is performed [1], owing to which the order again returns to the observed quantity — the single quantum longitudinal magnetization. The amplitude of the partial (for a given value of $M$) magnetization can be measured by means of a $\pi/2$ pulse that turns the magnetization into the plane perpendicular to the external magnetic field. Then one has to do Fourier transformation with respect to $\varphi$. To determine the relaxation rate, one repeats the experiment many times for different values of $t$.

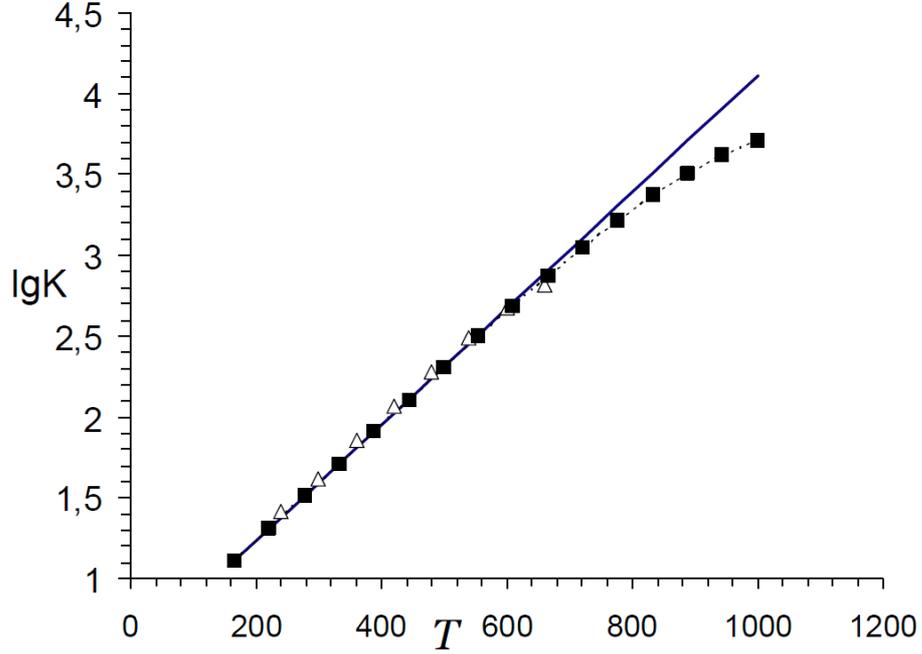

Fig. 2. Size $K$ of the effective cluster (decimal logarithm) as a function of the pumping time $T$ ($\mu s$) in adamantane. Triangles are the experimental results [8]. Black squares are experimental results [9]. Solid line is computation via Eq. (5). Dashed line is the result of approximation using relaxation processes [12].

Then, for the experimental method Fig. 1(a), the MQ coherence profile as a function of the order $M$ and time is expressed as

$$f_{0M}(T,t) = \frac{2}{\sqrt{\pi K(T)}} \exp\left\{-\frac{M^2}{K(T)}\right\} \exp\{-A^2 M^2 t^2\} \exp\{-K(T)b^2 t^2/2\}. \tag{7}$$

The constants $A^2$ and $b^2$ are directly related to the lattice sums of the coefficients $b_{ij}$ of Hamiltonian (1). From the experimental results in adamantane [8] in Ref. [20] we extract value $A^2 \approx 200$ $(1/ms)^2$ for $K=650$ spins (see Fig. 3).

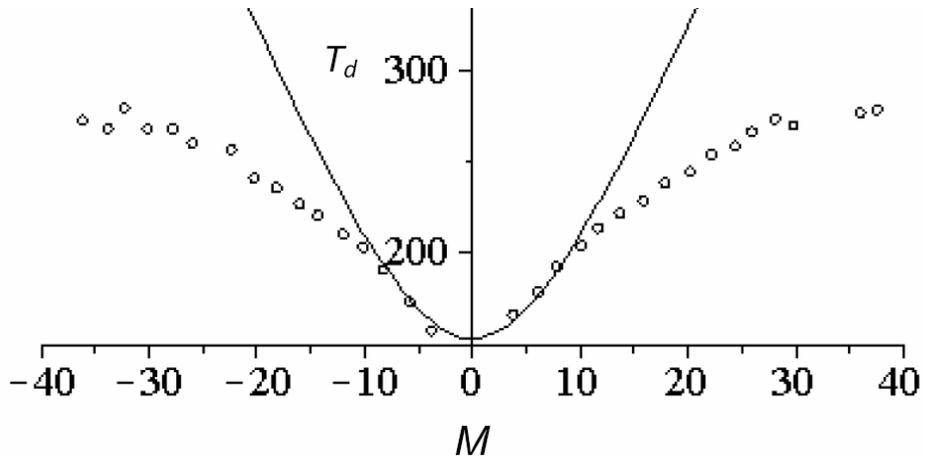

Fig. 3. Decoherence rates $1/T_d$ (1/ms) for different coherence orders for cluster sizes $K=650$ ($T = 660\mu s$). Circles are experimental rates [8] and line is $1/T_d = 205{,}48M^2 + 23145{,}1$.

As in the experiments in [8-10], the authors determine the average effective size $K_{eff}(T,t)$ of a cluster by the half-maximum bandwidth of the spectrum; therefore, it is expedient to introduce again the effective Gaussian distribution function that characterizes the intensity profile as a function of $M$:

$$\exp\left\{-\frac{M^2}{K_{eff}(T,t)}\right\} = \exp\left\{-\frac{M^2}{K(T)}\right\}\exp\{-A^2M^2t^2\}.$$

Now, the degradation of a cluster defined by time dependent Gaussian multipliers in Eq. (7) can be taken into consideration as follows:

$$K_{eff}(T,t) = \frac{1}{1/K(T) + A^2t^2}. \qquad (8).$$

The first and the last multipliers in Eq. (7) have no effect on the effective size of a cluster because they do not depend on the number $M$. The effect of these multipliers manifests itself in the variation of the intensity of the entire spectrum. According to Eq. (8), the bandwidth of the MQ spectrum and the $K_{eff}(T,t)$ decrease as the decay time increases (the evolution time with Hamiltonian (1)), which completely agrees with the experimental results of Ref.[9].

## 3. GROWTH AND DECAY OF MQ COHERENCES IN THE SIMULTANEOUS PROCESS OF FORMATION AND DEGRADATION OF CORRELATIONS IN A SPIN SYSTEM

Unlike conventional MQ experiments, the variations that were introduced in Refs. [13-15] (see Fig. 1(b)) lead to the substitution of a single Hamiltonian by the new special effective Hamiltonian

$$H_{eff} = (1-p)H_0 + pH_d, \qquad (9)$$

constructed from Hamiltonians (1) and (2). In previous scheme both of Hamiltonians act separately at different time intervals. And now we have

$$p = \tau_1/\tau_c, \quad \tau_c = \tau_0 + \tau_1.$$

As is shown below, these variations manifest themselves in the experiment, first, in the decrease of the growth rate of MQ coherences compared with the case of $p = 0$ and, second, to change the damping, compared with the decay given by equation (7).

Since a quantity that characterizes the number of spins between which a dynamic correlation is established during the preparatory period is given by the second moment of the MQ coherence intensity profile $K(N\tau_c) = 2 << M^2 >>$ (see Refs. [11, 21]), we should next to evaluate the variation in the time dependence of this second moment for $0 < p < 1$ compared with the result given by (5). In Ref. [18] for $p << 1$ we obtained the estimation

$$a_p \approx a_0(1-p) \qquad (10)$$

for the exponent

$$K(T) = \exp(a_p T), \qquad (11)$$

what describes the increasing of the second moment with time $T = N\tau_c$.

If $p<<1$ in the case of simultaneous emergence and degradation of MQ coherences (see Fig. 1(b)), a coherence that arises at time instant $t$ under the interaction $H_0$ from Hamiltonian (9) on the time interval [0, $T$] will further degrade under the interaction $pH_d$ from Hamiltonian (9). Then the decay occurs during a time interval of $T - t$. As it follows from the aforesaid, to describe the degradation in the experiment (Fig. 1(b)), one has to replace the instant time $t$ by average with respect to the emergence instant of a coherence. Thus for a function describing the decay of a coherence with a given $M$, we obtain:

$$<\Gamma_M(t)>_t = <\exp\{-p^2A^2M^2(T-t)^2\}>. \qquad (12)$$

To perform the averaging in Eq. (12), we should find $R(t)$, the probability density of emerging a coherence as a function of time. The sought probability density is determined by the time derivative of expression (11):

$$R(t) = \frac{1}{D}\frac{K(t)}{dt} = \frac{a_p}{D}\exp(a_p t) \quad (13)$$

where $D$ is a normalization constant,

$$D = \exp(a_p T) - 1.$$

Now we can obtain the time average in (12) and we can find the sought result for the TCF describing the relaxation of coherence intensities of different orders in the MQ spectrum:

$$<\Gamma_M(t)>_t = \int_0^T \exp\{-p^2 A^2 M^2 (T-t)^2\} R(t) dt = \frac{U_2(y,m)}{1-e^{-y}}, \quad (14)$$

where $y = a_p T$, $m = |M|\frac{Ap}{a_p}$,

$$U_2(y,m) = \int_0^y e^{-x} e^{-(mx)^2} dx = \frac{\sqrt{\pi}}{2m} \exp\left(\frac{1}{4m^2}\right)\left[erf\left(ym + \frac{1}{2m}\right) - erf\left(\frac{1}{2m}\right)\right]. \quad (15)$$

Then the MQ spectrum with the relaxation is

$$f_M(T) \propto g_M(T) <\Gamma_M(t)>_t. \quad (16)$$

Here we will take for the calculations an initial spectrum in the next form

$$g_M(T) \propto \exp\left\{-\left(\frac{M^2}{K(T)}\right)^{\lambda/2}\right\}. \quad (17)$$

If $\lambda = 2$ the above expression represents the Gaussian function and whether $\lambda = 1$ it is the exponential one. Let's note that the authors of Ref. [14] argue that the distribution of coherences of different orders in the MQ spectrum preferably to describe by the exponential form of function from Eq. (17) rather than by the Gaussian one. Now we should determine the average effective cluster size using the relation $K_{eff} = M_e^2$, and so through the order of coherence $M_e$, in situation when the intensity of the MQ spectrum is reduced by a factor $e$. By this way to find $M_e$ we obtain the equation

$$\exp\left\{-\left(\frac{M_e^2}{K(\tau)}\right)^{\lambda/2}\right\}\frac{U_2(y,m_e)}{1-e^{-y}} = \frac{1}{e}. \quad (18)$$

where $m_e = |M_e|\frac{Ap}{a_p}$. At $y = a_p T << 1$ we have

$$M_e \approx e^{y/2}\left\{1 - \frac{1}{3\lambda}\left(\frac{yAp}{a_p}e^{y/2}\right)^2\right\}.$$

At $y = a_p T >> 1$ $K_{eff}$ reaches its steady-state value

$$K_{st} \approx \frac{3,2 a_p^2}{A^2 p^2}. \quad (19)$$

Equation (18) has been solved numerically and needed values of variable by experimentalists parameter $p$ where taken from the article [14]. The results obtained are shown in Fig. 4.

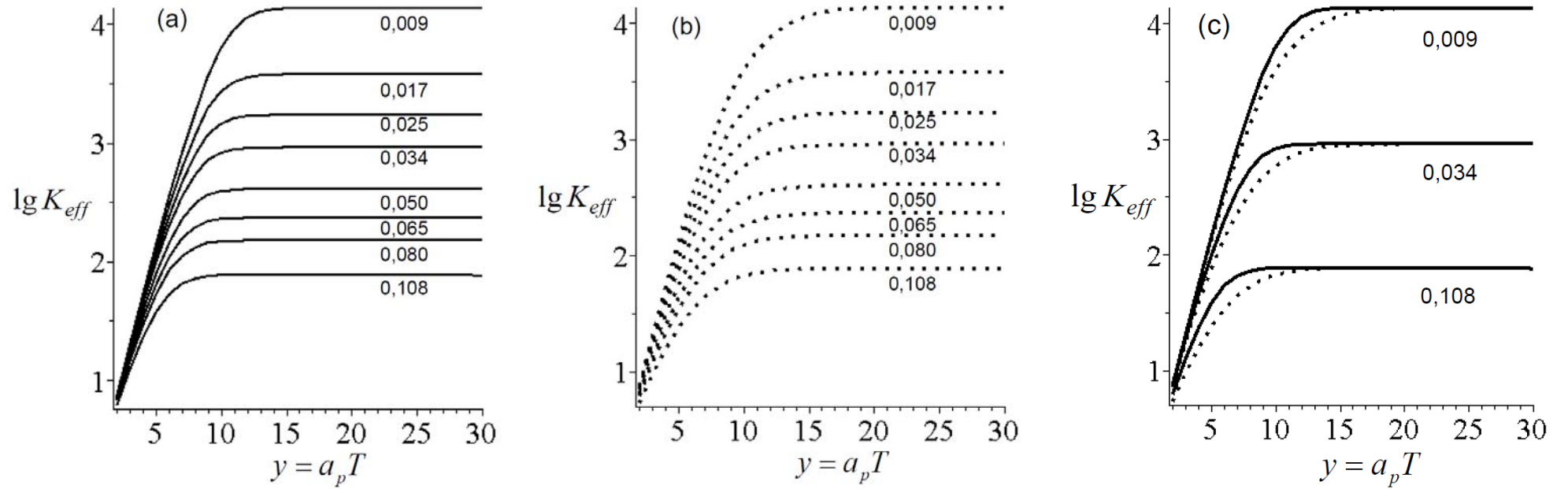

**Fig. 4.** Evolution of the effective cluster size under conditions displayed at Fig. 1(b) for various values of the parameter $p$ (the numbers under appropriate curves). (a) curves for the Gaussian form of MQ spectrum (17), and (b) curves for the exponential form. (c) curves for both forms on the same graph (for the Gaussian form shown by solid lines and for the exponential form shown by dotted lines) The abscissa shows the time in units of $1/a_p$

## 4. DISCUSSION

The time dependence of the effective number $K_{eff}$ of correlated spins illustrated at Fig. 4 shows good agreement with the experimental results of Ref. [14] and adequately reflects all of the characteristic features of behavior of the experimental functions. We should mark special emphasis on the most important result of the theory presented here. The fact that the effective cluster size $K_{eff}$ reaches a steady state value $K_{st}$ predicted by Eq. (19). Such a stabilization of the cluster size indeed was observed in Ref. [14]; in exact conformity with Eq. (19), the experimental effective cluster size was found inversely proportional to the square of the parameter $p$. Experimental and theoretical dependence $K_{st}$ vs. $p$ are shown in Fig. 5. The experimental value of the coefficient multiplying this function did not too adequate coincides with the ratio of constants from Eq. (19) what earlier been determined in Refs. [12, 20] (see also Fig. 2 and Fig. 3). As we suppose the principle reason for it is that the available time in the experiment ( $y \leq 10$ ) has not yet reached one needed for achievement the steady-state value (19). Especially it is true for small perturbations.

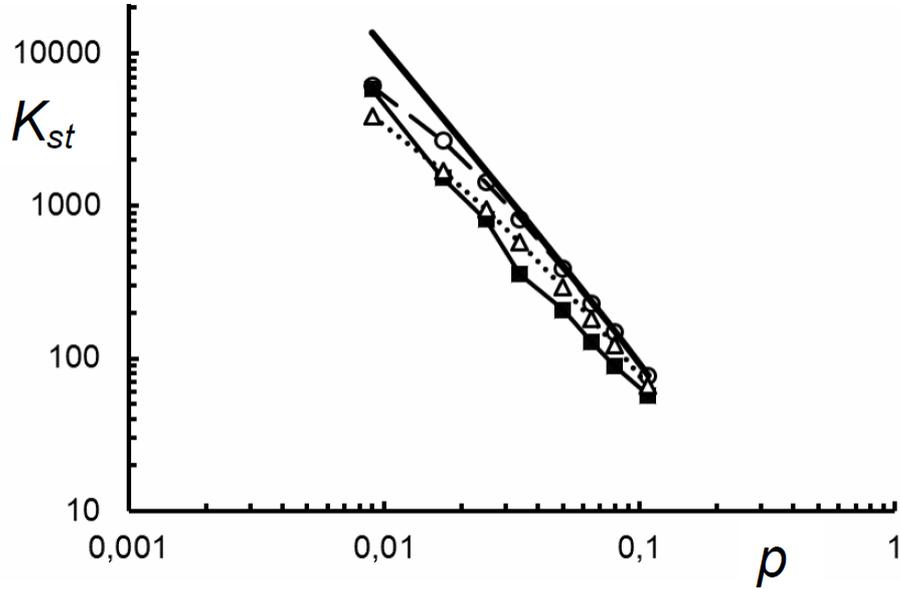

Fig. 5. The stabilized cluster size $K_{st}$ of correlated spins versus the perturbation strength $p$. The experimental results from Ref. [14] are shown by black squares. The results from Fig. 4 at $y$=10 are shown by circles for the Gaussian form of MQ spectrum (17) and by triangles for the exponential form. The results for both forms at $y$=30 witch equal (19) are shown by solid line.

According to the aforesaid, this stabilization of the cluster size is associated with the behavior of the MQ coherence profile as a function of the order $M$. Indeed, the dependence of the MQ spectrum on $M$ is expressed as

$$f_M(T) = g_M(T) < \Gamma_M(t) >_t \propto \exp\left\{-\left(\frac{M^2}{K(T)}\right)^{z/2}\right\} < \exp\{-p^2 A^2 M^2 (T-t)^2\} >. \quad (20)$$

The number $K$ in Eq. (20) grows exponentially with $T$; therefore, the dependence on $M$ of the first cofactor becomes weaker, and the dependence of the MQ spectrum on $M$ is completely determined by the second cofactor in (20). The maximum number of coherences arises near the boundary $t = T$. At a distance of $\Delta t$ away from this boundary toward smaller times, the number of emerging coherences decays exponentially ($R(t) \sim \exp(-a_p \Delta t)$), so when Eqs. (12) – (14) are averaged the main contribution comes from the area $\Delta t \sim 1/a_p \ll T$ and it becomes independent of *T*. Thus, the intensity profile becomes time independent; therefore, the mean cluster size defined by the half-maximum bandwidth also becomes independent of *T*, whereas in reality cluster continues to grow.

On the other hand, we should mark the contradiction for the interpretation given at Refs. [13-15, 17] where authors used the conception of "localization". Anyone can points out the following arguments. Let's suppose that we are switch on the interaction $pH_d$ only at the last stage of experiment: i.e. at the stage of reverse evolution while the preparation of correlations we are doing with the ideal Hamiltonian -$H_0$. For the theory, this changing of the plan from the first one means that one has to do cyclic permutation of corresponding operators of evolution in the trace what describes the expression for observable signal,

$$\Gamma_\varphi(N\tau_0, N\tau_c) = Tr\{U_0^+ U_{\varphi z} U_p S_z U_p^+ U_{\varphi z}^+ U_o S_z\} / Tr\{S_z^2\},$$

were $U_p = \exp\{-iN\tau_c(1-p)H_0 - iN\tau_c pH_d\}$, $U_0 = \exp\{-iN\tau_0 H_0\}$, $U_{\varphi z}=exp(i\varphi S_z)$ is operator of rotation through an angle $\varphi$ around the z axis. Certainly, the above procedure from the mathematical point of view means only identical transformation but now, after the permutation, there are no any reasons to state something about "localization" at the preparation stage.

There is why one can see that stabilization of width observed experimentally in Refs. [13-15] probably is the consequence of decreasing of MQ spectral components and there are no any reasons for extinction of growing up the effective cluster size.

# REFERENCES.

1. J. Baum, M. Munovitz, A.N. Garroway, A. Pines, J. Chem. Phys. **83**, 2015 (1985).
2. R.E. Ernst, G. Bodenhausen, and A. Wokaun, *Principles of Nuclear Magnetic Resonance in One and Two Dimensions*, Oxford Univ. Press, Oxford (1987)
3. M. Munovitz, A. Pines, Adv. Chem. Phys. **6**, 1 (1987).
4. P.-K. Wang, J.-P. Ansermet, S.L. Rudaz, Z. Wang, S. Shore, C.P. Slichter, J.H. Sinfelt, Science **234**, 35 (1986).
5. J. Baum, A. Pines, J. Am. Chem. Soc. **108**, 7447 (1986).
6. S.I. Doronin, A.V. Fedorova, E.B. Fel'dman, A.I. Zenchuk, J. Chem. Phys. **131**, 104109 (2009).
7. J.A. Jones, *Quantum Computing with NMR*, Prog. NMR Specrosc. **59**, 91 (2011).
8. H.G. Krojanski, D. Suter, Phys. Rev. Lett. **93**, 090501 (2004).
9. H.G. Krojanski, D. Suter, Phys. Rev. A **74**, 062319 (2006).
10. G. Cho, P. Cappelaro, D.G. Cory, C. Ramanathan, Phys. Rev. B **74**, 224434 (2006).
11. V.E. Zobov, A.A. Lundin, JETP **103**, 904 (2006).
12. V.E. Zobov, A.A. Lundin, Russian Journal of Physical Chemistry B **2,** 676 (2008).
13. G.A. Alvarez, D. Suter, Phys. Rev. Lett. **104**, 230403 (2010).
14. G.A. Alvarez, D. Suter, Phys. Rev. A **84**, 012320 (2011).
15. G.A. Alvarez, R. Kaiser, D. Suter, Ann. Phys. (Berlin) **525**, 833 (2013).
16. V.L. Bodneva, A.A. Lundin, JETP **116**, 1050 (2013).
17. G.A. Alvarez, D. Suter, R. Kaiser, arXiv:1409.4562 (2014).
18. V.E. Zobov, A.A. Lundin, JETP **113**, 1006, (2011).
19. A. Abragam, *Principles of Nuclear Magnetism*, Oxford Univ. Press, Oxford, (1961).
20. V.E. Zobov, A.A. Lundin, JETP **112**, 451 (2011).
21. A.K. Khitrin, Chem. Phys. Lett. **274**, 217 (1997).